\pdfoutput=1

\documentclass[12pt,a4paper]{article}

\usepackage{ifthen} 
\newboolean{pdflatex}
\setboolean{pdflatex}{true} 

\newboolean{articletitles}
\setboolean{articletitles}{true} 

\newboolean{uprightparticles}
\setboolean{uprightparticles}{false} 

\usepackage{multirow}
\usepackage{booktabs}
\usepackage{longtable}
\usepackage{amsmath}
\usepackage{verbatim}
\usepackage{epstopdf}

\usepackage{microtype}
\usepackage{lineno}  
\usepackage{xspace} 

\usepackage{graphicx}  
\usepackage{color}
\usepackage{colortbl}
\graphicspath{{./figs/}} 

\usepackage{amsmath} 
\usepackage{amssymb}
\usepackage{amsfonts}
\usepackage{upgreek} 

\newcommand*\patchAmsMathEnvironmentForLineno[1]{%
\expandafter\let\csname old#1\expandafter\endcsname\csname #1\endcsname
\expandafter\let\csname oldend#1\expandafter\endcsname\csname
end#1\endcsname
 \renewenvironment{#1}%
   {\linenomath\csname old#1\endcsname}%
   {\csname oldend#1\endcsname\endlinenomath}%
}
\newcommand*\patchBothAmsMathEnvironmentsForLineno[1]{%
  \patchAmsMathEnvironmentForLineno{#1}%
  \patchAmsMathEnvironmentForLineno{#1*}%
}
\AtBeginDocument{%
\patchBothAmsMathEnvironmentsForLineno{equation}%
\patchBothAmsMathEnvironmentsForLineno{align}%
\patchBothAmsMathEnvironmentsForLineno{flalign}%
\patchBothAmsMathEnvironmentsForLineno{alignat}%
\patchBothAmsMathEnvironmentsForLineno{gather}%
\patchBothAmsMathEnvironmentsForLineno{multline}%
}

\usepackage{hyperref}    
\usepackage[all]{hypcap} 

\def\lhcb {\mbox{LHCb}\xspace}
\def\ux85 {\mbox{UX85}\xspace}

\ifthenelse{\boolean{uprightparticles}}%
{

 \def\Ppi         {\ensuremath{\uppi}\xspace}

 \def\Ppsi        {\ensuremath{\uppsi}\xspace}

 \def\PDelta      {\ensuremath{\Delta}\xspace}                 
 \def\PXi      {\ensuremath{\Xi}\xspace}                 
 \def\PLambda      {\ensuremath{\Lambda}\xspace}                 
 \def\PSigma      {\ensuremath{\Sigma}\xspace}                 
 \def\POmega      {\ensuremath{\Omega}\xspace}                 
 \def\PUpsilon      {\ensuremath{\Upsilon}\xspace}                 
 
 \def\PB      {\ensuremath{\mathrm{B}}\xspace}                 
                  
 \def\PD      {\ensuremath{\mathrm{D}}\xspace}

 \def\PJ      {\ensuremath{\mathrm{J}}\xspace}                 
 \def\PK      {\ensuremath{\mathrm{K}}\xspace}

 \def\Pb      {\ensuremath{\mathrm{b}}\xspace}                 
 \def\Pc      {\ensuremath{\mathrm{c}}\xspace}

 \def\Pi      {\ensuremath{\mathrm{i}}\xspace}

}
{

 \def\Ppi         {\ensuremath{\pi}\xspace}

 \def\Ppsi        {\ensuremath{\psi}\xspace}                 
                  
 \mathchardef\PDelta="7101
 \mathchardef\PXi="7104
 \mathchardef\PLambda="7103
 \mathchardef\PSigma="7106
 \mathchardef\POmega="710A
 \mathchardef\PUpsilon="7107
                  
 \def\PB      {\ensuremath{B}\xspace}                 
                  
 \def\PD      {\ensuremath{D}\xspace}

 \def\PJ      {\ensuremath{J}\xspace}                 
 \def\PK      {\ensuremath{K}\xspace}

 \def\Pb      {\ensuremath{b}\xspace}                 
 \def\Pc      {\ensuremath{c}\xspace}

 \def\Pi      {\ensuremath{i}\xspace}

}

\def\cquark    {\ensuremath{\Pc}\xspace}

\def\bquark    {\ensuremath{\Pb}\xspace}

\def\pion  {\ensuremath{\Ppi}\xspace}

\def\pip   {\ensuremath{\pion^+}\xspace}

\def\kaon  {\ensuremath{\PK}\xspace}
  \def\Kbar  {\kern 0.2em\overline{\kern -0.2em \PK}{}\xspace}

\def\Kz    {\ensuremath{\kaon^0}\xspace}
\def\Kzb   {\ensuremath{\Kbar^0}\xspace}
\def\KzKzb {\ensuremath{\Kz \kern -0.16em \Kzb}\xspace}
\def\Kp    {\ensuremath{\kaon^+}\xspace}
\def\Km    {\ensuremath{\kaon^-}\xspace}

\def\KpKm  {\ensuremath{\Kp \kern -0.16em \Km}\xspace}
\def\KS    {\ensuremath{\kaon^0_{\rm\scriptscriptstyle S}}\xspace}

  \def\Dbar    {\kern 0.2em\overline{\kern -0.2em \PD}{}\xspace}
\def\D       {\ensuremath{\PD}\xspace}

\def\Dz      {\ensuremath{\D^0}\xspace}
\def\Dzb     {\ensuremath{\Dbar^0}\xspace}
\def\DzDzb   {\ensuremath{\Dz {\kern -0.16em \Dzb}}\xspace}
\def\Dp      {\ensuremath{\D^+}\xspace}
\def\Dm      {\ensuremath{\D^-}\xspace}

\def\DpDm    {\ensuremath{\Dp {\kern -0.16em \Dm}}\xspace}

\def\B       {\ensuremath{\PB}\xspace}
  \def\Bbar    {\kern 0.18em\overline{\kern -0.18em \PB}{}\xspace}

\def\Bu      {\ensuremath{\B^+}\xspace}

\def\Bc      {\ensuremath{\B_\cquark^+}\xspace}

\def\jpsi     {\ensuremath{{\PJ\mskip -3mu/\mskip -2mu\Ppsi\mskip 2mu}}\xspace}

  \def\Y#1S{\ensuremath{\PUpsilon{(#1S)}}\xspace}

\def\Lbar {\ensuremath{\kern 0.1em\overline{\kern -0.1em\PLambda}}\xspace}

\def\BF         {{\ensuremath{\cal B}\xspace}}

\def\BR         {\BF}

\def\to                 {\ensuremath{\rightarrow}\xspace}

\def\AT#1     {\ensuremath{A_{\mathrm{T}}^{#1}}\xspace}           

\def\C#1      {\ensuremath{\mathcal{C}_{#1}}\xspace}                       
\def\Cp#1     {\ensuremath{\mathcal{C}_{#1}^{'}}\xspace}                    
\def\Ceff#1   {\ensuremath{\mathcal{C}_{#1}^{\mathrm{(eff)}}}\xspace}        
\def\Cpeff#1  {\ensuremath{\mathcal{C}_{#1}^{'\mathrm{(eff)}}}\xspace}       
\def\Ope#1    {\ensuremath{\mathcal{O}_{#1}}\xspace}                       
\def\Opep#1   {\ensuremath{\mathcal{O}_{#1}^{'}}\xspace}                    

\newcommand{\tev}{\ensuremath{\mathrm{\,Te\kern -0.1em V}}\xspace}
\newcommand{\gev}{\ensuremath{\mathrm{\,Ge\kern -0.1em V}}\xspace}
\newcommand{\mev}{\ensuremath{\mathrm{\,Me\kern -0.1em V}}\xspace}
\newcommand{\kev}{\ensuremath{\mathrm{\,ke\kern -0.1em V}}\xspace}
\newcommand{\ev}{\ensuremath{\mathrm{\,e\kern -0.1em V}}\xspace}
\newcommand{\gevc}{\ensuremath{{\mathrm{\,Ge\kern -0.1em V\!/}c}}\xspace}
\newcommand{\mevc}{\ensuremath{{\mathrm{\,Me\kern -0.1em V\!/}c}}\xspace}
\newcommand{\gevcc}{\ensuremath{{\mathrm{\,Ge\kern -0.1em V\!/}c^2}}\xspace}
\newcommand{\gevgevcccc}{\ensuremath{{\mathrm{\,Ge\kern -0.1em V^2\!/}c^4}}\xspace}
\newcommand{\mevcc}{\ensuremath{{\mathrm{\,Me\kern -0.1em V\!/}c^2}}\xspace}

\def\mum  {\ensuremath{\,\upmu\rm m}\xspace}

\def\invfb   {\ensuremath{\mbox{\,fb}^{-1}}\xspace}

\def\gsim{{~\raise.15em\hbox{$>$}\kern-.85em
          \lower.35em\hbox{$\sim$}~}\xspace}
\def\lsim{{~\raise.15em\hbox{$<$}\kern-.85em
          \lower.35em\hbox{$\sim$}~}\xspace}

\def\sPlot{\mbox{\em sPlot}}

\def\sqs   {\ensuremath{\protect\sqrt{s}}\xspace}

\def\evtgen     {\mbox{\textsc{EvtGen}}\xspace}
\def\pythia     {\mbox{\textsc{Pythia}}\xspace}

\def\geant      {\mbox{\textsc{Geant4}}\xspace}

\def\photos     {\mbox{\textsc{Photos}}\xspace}

\def\tell1  {TELL1\xspace}
\def\ukl1   {UKL1\xspace}

\newcommand{\pd}[1]{${#1}$}

\newcommand{\ptrans}{\ensuremath{p_{\rm T}}}

\newcommand{\bsubc}{\ensuremath{B_c^+}}
\newcommand{\bplus}{\ensuremath{B^+}}
\newcommand{\bcjpsipi}{\ensuremath{\bsubc\to\jpsi \pi^+}}

\newcommand{\bpjpsik}{\ensuremath{\bplus\to\jpsi K^+}}

\usepackage{mciteplus}

\newcommand{\BcMassResult}{\ensuremath{6273.7 \pm 1.3\,({\rm stat.})
    \pm 1.6 \,({\rm syst.}) \,\mevcc}}
\newcommand{\BcMassDiffResult}{\ensuremath{994.6 \pm 1.3\,({\rm stat.}) \pm 0.6\,({\rm syst.}) \,\mevcc}}
\newcommand{\BcCSResult}{\ensuremath{(0.68 \pm 0.10\,({\rm stat.}) \pm
    0.03\,({\rm syst.}) \pm 0.05\,({\rm lifetime}) )\%}}

\begin{document}

\renewcommand{\thefootnote}{\fnsymbol{footnote}}
\setcounter{footnote}{1}

\begin{titlepage}
\pagenumbering{roman}

\vspace*{-1.5cm}
\centerline{\large EUROPEAN ORGANIZATION FOR NUCLEAR RESEARCH (CERN)}
\vspace*{1.5cm}
\hspace*{-0.5cm}
\begin{tabular*}{\linewidth}{lc@{\extracolsep{\fill}}r}
\ifthenelse{\boolean{pdflatex}}
{\vspace*{-2.7cm}\mbox{\!\!\!\includegraphics[width=.14\textwidth]{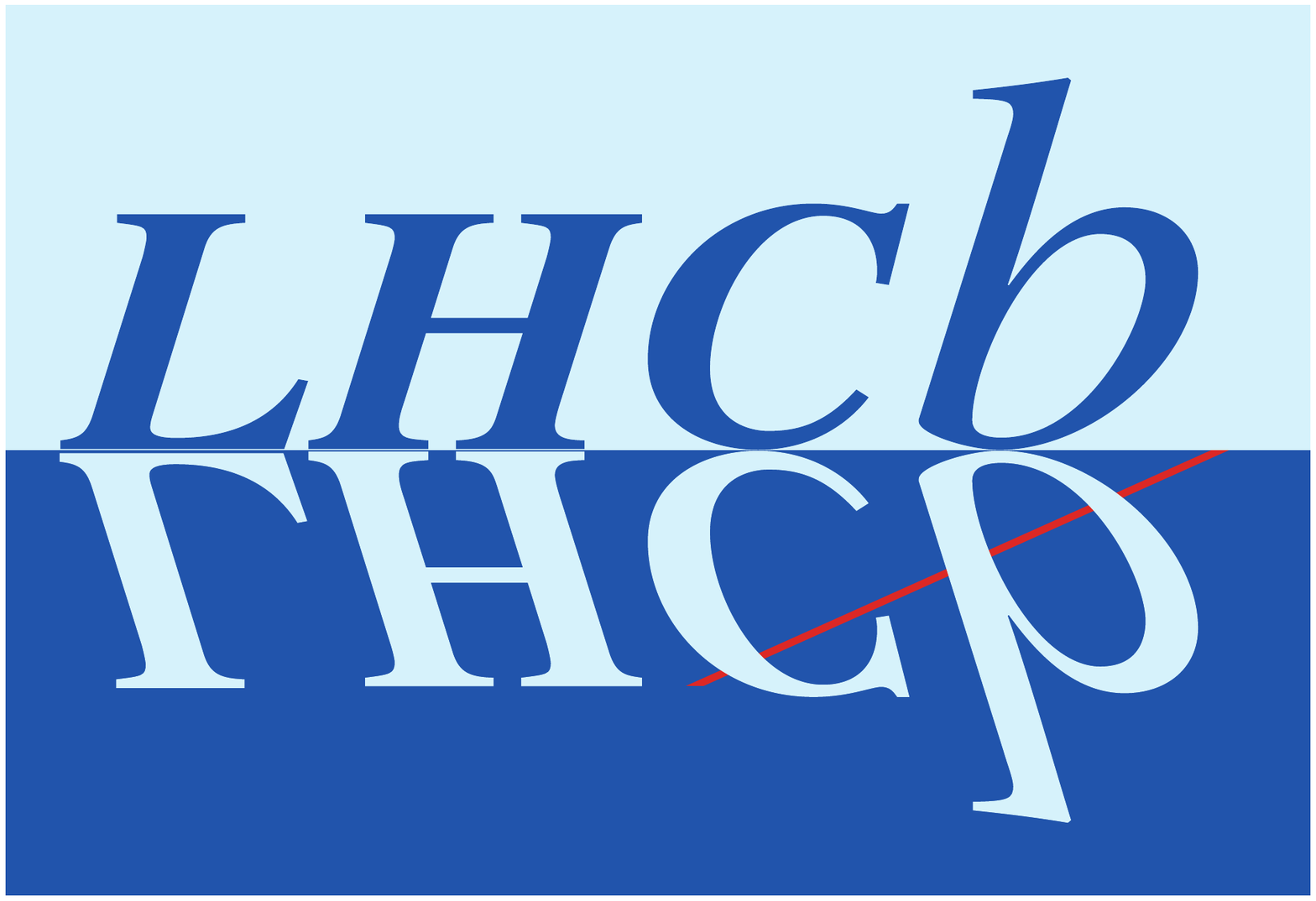}} & &}%
{\vspace*{-1.2cm}\mbox{\!\!\!\includegraphics[width=.12\textwidth]{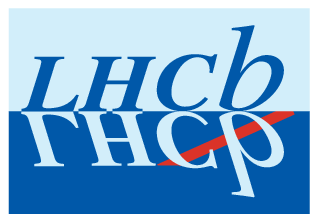}} & &}%
\\
& & CERN-PH-EP-2012-275 \\  
& & LHCb-PAPER-2012-028 \\  
& & October 18, 2012  \\  
& & \\
\end{tabular*}

\vspace*{4.0cm}

{\bf\boldmath\huge
\begin{center}
  Measurements of $B_c^+$ production and mass 
  with the $B_c^+ \to \jpsi \pi^+$ decay
\end{center}
}

\vspace*{1.5cm}

\begin{center}
The LHCb collaboration\footnote{Authors are listed on the following pages.}
\end{center}

\vspace{\fill}

\begin{abstract}
  \noindent
  Measurements of $B_c^+$ production and mass are
  performed with the decay mode
  $B_c^+ \to \jpsi \pi^+$ using 0.37 fb$^{-1}$ of data collected
  in $pp$ collisions at $\sqrt{s}=7$ TeV by the LHCb
  experiment. The ratio of the production cross-section times
  branching fraction between the 
  $B_c^+ \to \jpsi \pi^+$ and the 
  $B^+ \to \jpsi K^+$ decays is measured to be \BcCSResult\ 
  for $\Bc$ and $\Bu$ mesons with transverse momenta $p_{\rm T}>4~$GeV/$c$ 
  and pseudorapidities $2.5<\eta<4.5$.
  The $B_c^+$ mass is directly measured to be $\BcMassResult$,
  and the measured mass difference with respect to the $\Bu$ meson is $M(B_c^+)-M(B^+) = \BcMassDiffResult$.
\end{abstract}

\vspace*{1.5cm}

\begin{center}
  Submitted to Phys. Rev. Lett.
\end{center}

\vspace{\fill}

\end{titlepage}

\newpage
\setcounter{page}{2}
\mbox{~}
\newpage

\centerline{\large\bf LHCb collaboration}
\begin{flushleft}
\small
R.~Aaij$^{38}$, 
C.~Abellan~Beteta$^{33,n}$, 
A.~Adametz$^{11}$, 
B.~Adeva$^{34}$, 
M.~Adinolfi$^{43}$, 
C.~Adrover$^{6}$, 
A.~Affolder$^{49}$, 
Z.~Ajaltouni$^{5}$, 
J.~Albrecht$^{35}$, 
F.~Alessio$^{35}$
% ...

{\footnotesize \it
$ ^{1}$Centro Brasileiro de Pesquisas F\'{i}sicas (CBPF), Rio de Janeiro, Brazil\\
% ...
$ ^{5}$Clermont Universit\'{e}, Universit\'{e} Blaise Pascal, CNRS/IN2P3, LPC, Clermont-Ferrand, France\\
$ ^{6}$CPPM, Aix-Marseille Universit\'{e}, CNRS/IN2P3, Marseille, France\\
% ...
$ ^{11}$Physikalisches Institut, Ruprecht-Karls-Universit\"{a}t Heidelberg, Heidelberg, Germany\\
% ...
$ ^{33}$Universitat de Barcelona, Barcelona, Spain\\
$ ^{34}$Universidad de Santiago de Compostela, Santiago de Compostela, Spain\\
$ ^{35}$European Organization for Nuclear Research (CERN), Geneva, Switzerland\\
% ...
$ ^{38}$Nikhef National Institute for Subatomic Physics, Amsterdam, The Netherlands\\
% ...
$ ^{43}$H.H. Wills Physics Laboratory, University of Bristol, Bristol, United Kingdom\\
% ...
$ ^{49}$Oliver Lodge Laboratory, University of Liverpool, Liverpool, United Kingdom\\
% ...
$ ^{n}$LIFAELS, La Salle, Universitat Ramon Llull, Barcelona, Spain\\
% ...
}
\end{flushleft}

\cleardoublepage


\renewcommand{\thefootnote}{\arabic{footnote}}
\setcounter{footnote}{0}


\pagestyle{plain} 
\setcounter{page}{1}
\pagenumbering{arabic}

\noindent The $\Bc$ meson 
is unique in the Standard Model 
as it is the ground state of a family 
of mesons containing two different heavy flavour quarks.
At the 7 TeV LHC centre-of-mass energy, the most probable way to produce 
$B_c^{(*)+}$ mesons is through the $gg$-fusion process,
$gg\rightarrow B_c^{(*)+}+b+\bar{c}$~\cite{Brambilla:2004wf}. 
The production cross-section of the $\Bc$ meson
has been calculated by a complete order-$\alpha_s^4$ approach
and using the fragmentation approach~\cite{Brambilla:2004wf}.
It is predicted to be about 0.4 $\upmu$b~\cite{Chang:2003cr, Gao:2010zzc} at $\sqrt{s}=7$~TeV
including contributions from excited states. 
This is one order of magnitude higher than that predicted at
the Tevatron energy $\sqrt{s}=1.96$~TeV. 
However, the theoretical predictions suffer from large uncertainties, 
and an accurate measurement of the $\Bc$ production cross-section 
is needed to guide experimental studies at the LHC. 
As is the case for heavy quarkonia, the mass of the $\Bc$ meson can
be calculated by means of potential models and lattice QCD, 
and early predictions lay in
the range from $6.2 - 6.4\,\gevcc$~\cite{Brambilla:2004wf}.
The inclusion of charge conjugate modes is implied throughout this Letter.

The $\Bc$ meson was first observed
in the semileptonic decay mode
$B_c^+ \rightarrow \jpsi(\mu^+\mu^-) \ell^+ X\ (\ell=e,\mu)$
by CDF \cite{Abe:1998wi, *Abe:1998fb}. 
The production cross-section times 
branching fraction for 
this decay
relative to that for $B^+ \rightarrow \jpsi \, K^+$ was measured to be
$0.132\, ^{+0.041}_{-0.037} \, ({\rm stat.}) \, 
\pm 0.031 \, ({\rm syst.}) \,
^{+0.032}_{-0.020} \, ({\rm lifetime})$ 
for $\Bc$ and $\Bu$ mesons with transverse momenta $p_{\rm T}>6$ GeV/$c$ 
and rapidities $|y|<1$.
Measurements of the \Bc\ mass by CDF~\cite{Aaltonen:2007gv} and 
D0~\cite{Abazov:2008kv} using 
the fully reconstructed decay
\pd{B_c^+\rightarrow \jpsi(\mu^+\mu^-) \pi^+} 
gave $M(B_c^+)=6275.6 \pm 2.9\, ({\rm stat.}) \pm 2.5\,
({\rm syst.})\,\mevcc$
and 
$M(B_c^+)= 6300 \pm 14\, ({\rm stat.}) \pm 5\, ({\rm syst.})\,\mevcc$, respectively.
A more precise measurement of the $\Bc$ mass would allow for more stringent tests of
predictions from potential models and lattice QCD calculations.

In this Letter, we present a measurement of the ratio of the 
production cross-section times branching fraction
of $\Bc \to \jpsi \pi^+$ relative to that for $\Bu \to \jpsi K^+$ 
for $\Bc$ and $\Bu$ mesons 
with transverse momenta $p_{\rm T}>4$ GeV/$c$ 
and pseudorapidities $2.5<\eta<4.5$,
and a measurement of the $\Bc$ mass.
These measurements are performed 
using $0.37\,\invfb$ of data collected
in $pp$ collisions at $\sqrt{s}=7$ TeV by the LHCb experiment.
The \lhcb detector~\cite{Alves:2008zz} is a single-arm forward
spectrometer covering the pseudorapidity range $2<\eta <5$, designed
for the study of particles containing \bquark or \cquark quarks. The
detector includes a high precision tracking system consisting of a
silicon-strip vertex detector surrounding the $pp$ interaction region,
a large-area silicon-strip detector located upstream of a dipole
magnet with a bending power of about $4{\rm\,Tm}$, and three stations
of silicon-strip detectors and straw drift-tubes placed
downstream. The combined tracking system has a momentum resolution
$\Delta p/p$ that varies from 0.4\% at 5\gevc to 0.6\% at 100\gevc,
and an impact parameter (IP) resolution of 20\mum for tracks with high
transverse momentum. Charged hadrons are identified using two
ring-imaging Cherenkov detectors. Photon, electron and hadron
candidates are identified by a calorimeter system consisting of
scintillating-pad and pre-shower detectors, an electromagnetic
calorimeter and a hadronic calorimeter. Muons are identified by a muon
system composed of alternating layers of iron and multiwire
proportional chambers. The muon identification efficiency is about 97\%, 
with a misidentification probability $\epsilon(\pi \to \mu) \sim 3$\%.

The $\bcjpsipi$ and $\bpjpsik$ decay modes are topologically
identical and are selected with requirements as similar as possible to
each other.
Events are selected by a trigger system consisting of
a hardware stage, based on information 
from the calorimeter and muon systems, followed by a software stage 
which applies a full event reconstruction. 
At the hardware trigger stage, events are selected 
by requiring a single muon candidate or a pair of muon candidates 
with high transverse momenta.
At the software trigger stage~\cite{LHCb-PUB-2011-017,
  LHCb-PUB-2011-016}, 
events are selected 
by requiring a pair of muon candidates 
with invariant mass within $120\,\mevcc$ of the $\jpsi$ mass~\cite{PDG2012}, 
or a two- or three-track secondary 
vertex with a large track $\ptrans$ sum, 
a significant displacement from the primary interaction,
and at least one track identified as a muon. 

At the offline selection stage, 
$\jpsi$ candidates are formed from pairs of oppositely
charged tracks with transverse momenta $\ptrans>0.9\,\gevc$
and identified as muons.
The two muons
are required to originate from a common vertex.
Candidates with a dimuon invariant mass 
between $3.04\gevcc$ and $3.14\gevcc$ are combined 
with charged hadrons with $\ptrans>1.5\, \gevc$
to form the $\Bc$ and $\Bu$ meson candidates. 
The $\jpsi$ mass window is about seven times larger than the mass resolution.
No particle identification is used in the selection of the hadrons.
To improve the $\Bc$ and $\Bu$ mass resolutions, 
the mass of the $\mu^+\mu^-$ pair is constrained to the $\jpsi$ mass~\cite{PDG2012}.
The $b$-hadron candidates are required to have $\ptrans>4\,\gevc$, 
decay time $t>0.25\,{\rm ps}$ and pseudorapidity in the range  
$2.5<\eta<4.5$. The fiducial region is chosen to be 
well inside the detector acceptance to have a reasonably flat
efficiency over the phase space.
To further suppress background to the $\Bc$ decay, the 
IP $\chi^2$ values of the $\jpsi$ and $\pip$ candidates with respect to
any primary vertex (PV) in the event are required to be larger
than 4 and 25, respectively. The IP $\chi^2$ is
defined as the difference between the $\chi^2$ of the PV
reconstructed with and without the considered particle.
The IP $\chi^2$ of the $\Bc$ candidates with respect to at least one PV in the
event is required to be less than 25.
After all selection requirements are applied, no event 
has more than one candidate for the $\bcjpsipi$ decay, and less than
1\% of the events have more than one
candidate for the $\bpjpsik$ decay. Such multiple candidates are retained
and treated the same as other candidates; the associated
systematic uncertainty is negligible.

The ratio of the production cross-section times branching fraction measured in this analysis is 
\begin{eqnarray}
  \label{eq:CSRatioForm}
  \begin{aligned}
    R_{c/u} &=\frac{\sigma(\Bc) \, \BR (B_c^+ \rightarrow \jpsi \pi^+)}
    {\sigma(\Bu) \, \BR(B^+ \rightarrow \jpsi K^+)} \\
    &= \frac{N\left(\bcjpsipi\right)}{\epsilon_{\rm tot}^{c}}
    \frac{\epsilon_{\rm tot}^{u}}{N\left(\bpjpsik\right)},
  \end{aligned}
\end{eqnarray}
where $\sigma(\Bc)$ and $\sigma(\Bu)$ are the inclusive production
cross-sections of the $\Bc$ and $\Bu$ mesons in $pp$ collisions 
at $\sqs=7$ TeV,   
$\BR (B_c^+ \rightarrow \jpsi \pi^+)$ and  
$\BR(B^+ \rightarrow \jpsi K^+)$ are the branching fractions of
the reconstructed decay chains,
$N\left(\bcjpsipi\right)$ and $N\left(\bpjpsik\right)$ 
are the yields of
the $\bcjpsipi$ and $\bpjpsik$ signal decays,
and $\epsilon_{\rm tot}^{c}$, $\epsilon_{\rm tot}^{u}$ 
are the total efficiencies, 
including geometrical acceptance, reconstruction, selection and
trigger effects. 

\begin{figure}
  \centering
  \includegraphics[width=0.495\textwidth]{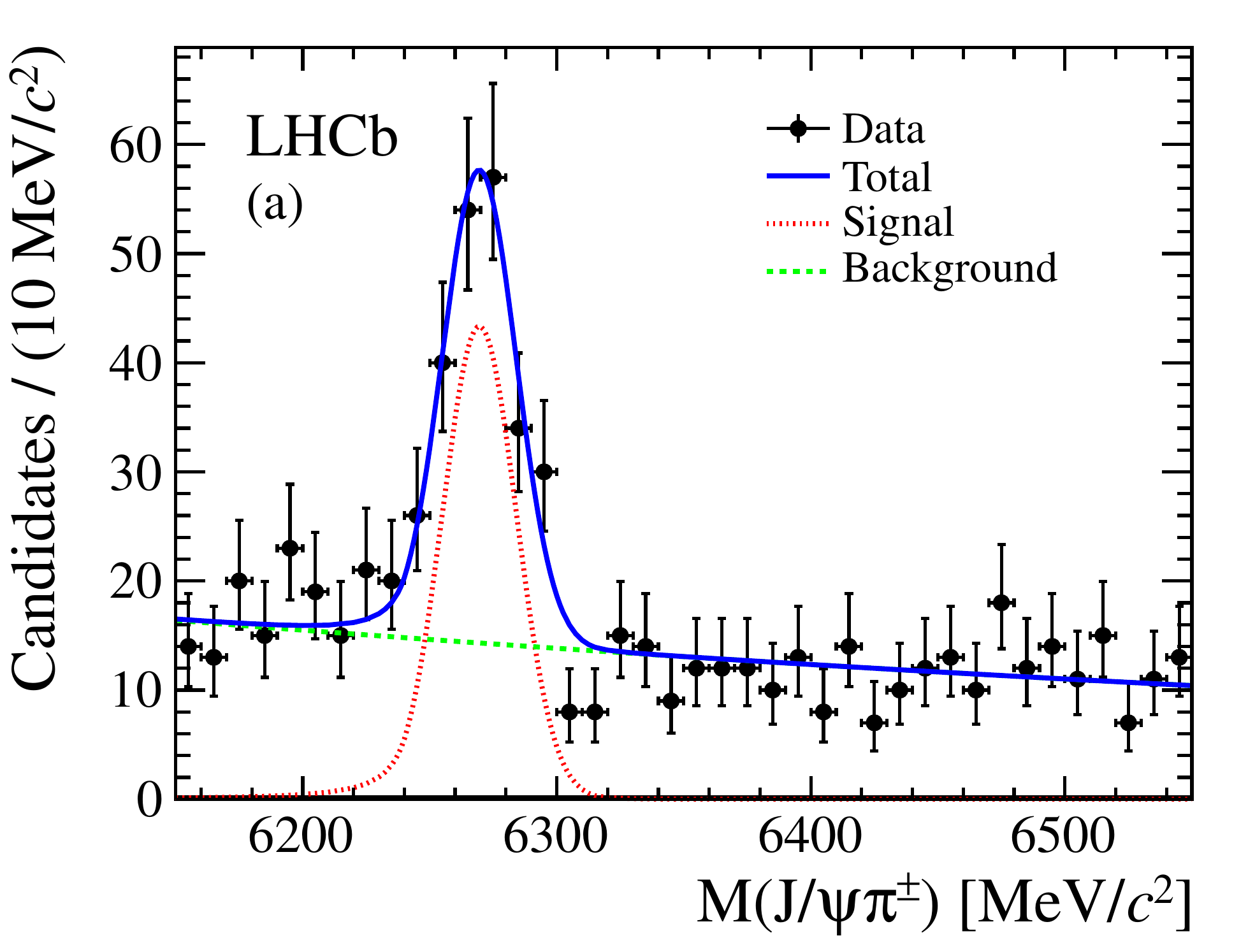}
  \includegraphics[width=0.495\textwidth]{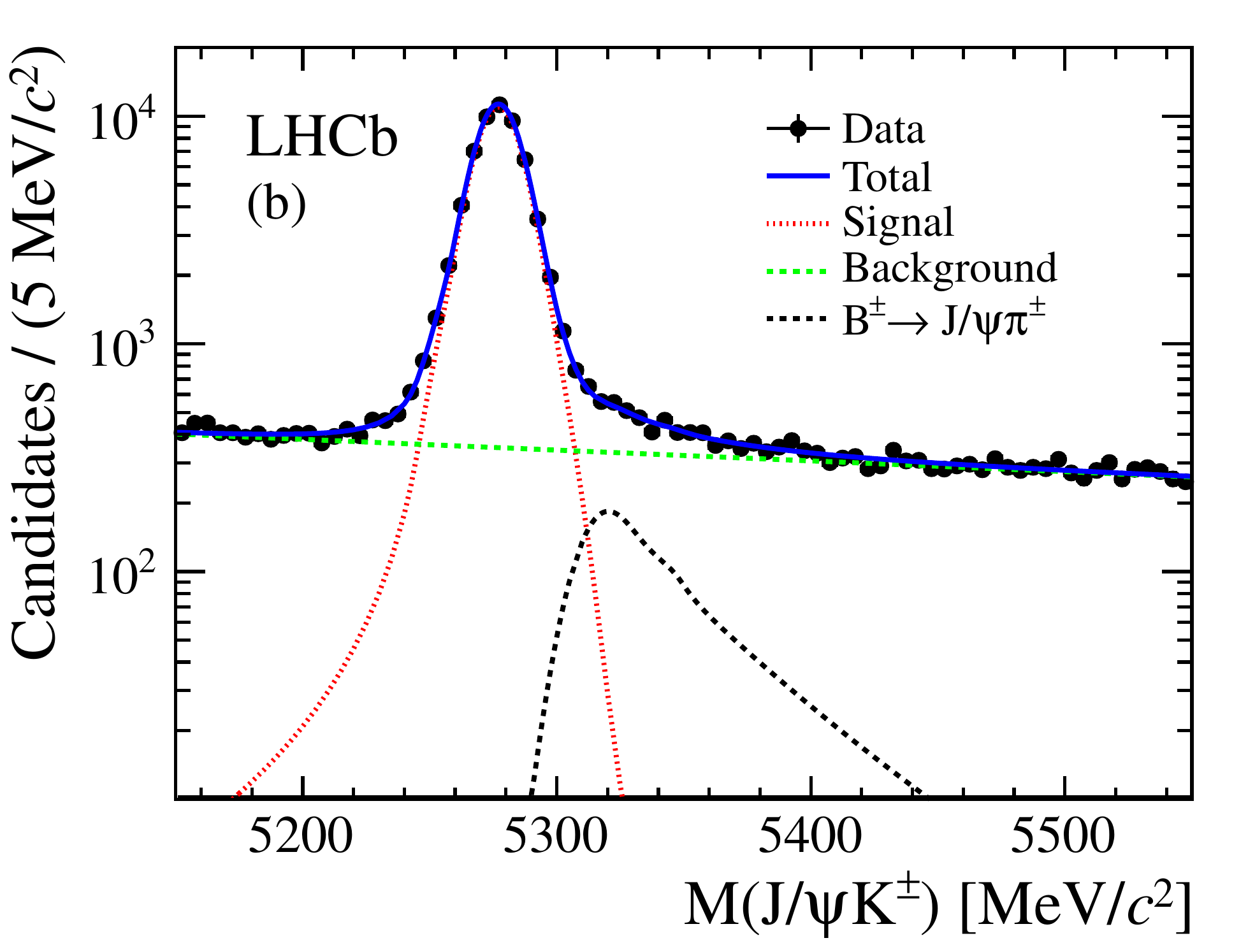}
  \caption{\small Invariant mass distributions of selected (a) $\bcjpsipi$ candidates
    and (b) $\bpjpsik$ candidates, used in the production measurement.
    The fits to the data are superimposed.} 
  \label{fig:ProdMassPlot}
\end{figure}

The signal event yields are obtained from extended unbinned
maximum likelihood fits to the invariant mass distributions of the reconstructed \Bc\ and
$\Bu$ candidates in the interval $6.15<M(\jpsi\pi^+)<6.55\gevcc$ for
$\Bc$ candidates and $5.15< M(\jpsi K^+)<5.55\gevcc$ for $\Bu$ candidates. 
The $\bcjpsipi$ signal mass shape is described by 
a double-sided Crystal Ball function~\cite{Skwarnicki:1986xj}.
The power law behaviour toward low mass is due primarily to 
final state radiation (FSR) from the bachelor hadron, 
whereas the high mass tail is mainly due to FSR from the muons 
in combination with the $\jpsi$ mass constraint.
The $\bpjpsik$ signal mass shape is described by 
the sum of two double-sided Crystal Ball functions that share the same
mean but have different resolutions.
From simulated decays, it is found that the tail parameters of the double-sided Crystal Ball function
depend mildly on the mass resolution. 
This functional dependence is determined from simulation and
included in the mass fit.
The combinatorial background is described by an exponential function. 
Background to $\bpjpsik$ from the Cabibbo-suppressed decay
$\Bu\to\jpsi\pi^+$ is included to improve the fit quality. 
The distribution is determined from the simulated events. 
The ratio of the number of 
$B^+\to \jpsi \pi^+$ decays to that of the signal is fixed 
to $\mathcal{B}(B^+\to \jpsi \pi^+)/\mathcal{B}(B^+\to \jpsi K^+)=3.83\%$~\cite{Aaij:2012jw}.
The Cabibbo-suppressed decay $\Bc\to\jpsi K^+$ is neglected as a
source of background to the $\bcjpsipi$ decay.
The invariant mass distributions of the selected $\bcjpsipi$
and $\bpjpsik$ candidates and the fits to the data 
are shown in Fig.~\ref{fig:ProdMassPlot}. 
The numbers of signal events are 
$162 \pm 18$ for $\bcjpsipi$ and $56\,243 \pm 256$ for $\bpjpsik$,
as obtained from the fits. 
The goodness of fits is checked with a $\chi^2$ test, 
which returns a probability of 97\% for $\bcjpsipi$
and 87\% for $\bpjpsik$.

The efficiencies, including geometrical acceptance, reconstruction, selection 
and trigger effects are determined using simulated signal events.
The production of the $\Bu$ meson is simulated using
\pythia~6.4~\cite{Sjostrand:2006za} with the configuration described
in Ref.~\cite{LHCb-PROC-2010-056}.
A dedicated generator~\textsc{BcVegPy}~\cite{Chang:2005hq} 
is used to simulate the $\Bc$ meson production. 
Decays of $\Bc$, $\Bu$ and $\jpsi$ mesons 
are described by \evtgen~\cite{Lange:2001uf} in which final state
radiation is generated using \photos~\cite{Golonka:2005pn}. 
The decay products are traced through the detector by the \geant
package~\cite{Allison:2006ve, *Agostinelli:2002hh} as described in
Ref.~\cite{LHCb-PROC-2011-006}.
As the efficiencies depend on $p_{\rm T}$ and $\eta$, 
the efficiencies from the simulation are binned in these variables 
to avoid a bias. 
The signal yield in each bin is obtained 
from data by subtracting
the background contribution using the $\sPlot$
technique~\cite{Pivk:2004ty},
where the signal and background mass shapes are assumed to
be uncorrelated with 
$p_{\rm T}$ and $\eta$. 
The efficiency-corrected numbers of $\bcjpsipi$ and $\bpjpsik$ signal 
decays are 
$2470 \pm 350$ 
and 
$364\,188 \pm 2270$, respectively, corresponding to a ratio of
$R_{c/u}=(0.68 \pm 0.10)\%$, where the uncertainties are statistical only.

The systematic uncertainties related to the determination of 
the signal yields and efficiencies are described in the following. 
Concerning the former, studies of simulated events 
show that effects due to the fit model 
on the measured ratio $R_{c/u}$ can be as much as 1\%,
which is taken as systematic uncertainty.  
The uncertainties from the contamination due to 
the Cabibbo-suppressed decays are found to be negligible. 

The uncertainties on the determination of the efficiencies are
dominated by the knowledge of the $\Bc$ lifetime,
which has been measured by CDF~\cite{Abulencia:2006zu} and
D0~\cite{Abazov:2008rba} to give 
$\tau(\Bc)=0.453\pm0.041\,{\rm ps}$~\cite{PDG2012}.
The distributions of the $\bcjpsipi$ simulated events have been
reweighted after changing the $\Bc$ lifetime 
by one standard deviation around 
its mean value and the efficiencies are recomputed. 
The relative difference of 7.3\% between the recomputed efficiencies 
and the nominal values is taken as
a systematic uncertainty. 
Since the $\Bu$ lifetime is known more precisely, 
its contribution to the uncertainty is neglected. 

The effects of the trigger requirements have been evaluated by only using the events
triggered by the lifetime unbiased (di)muon lines, 
which is about 85\% of the total number of events.
Repeating the complete analysis, a ratio of $R_{c/u} = (0.65 \pm
0.10)\%$ is found, resulting in a systematic uncertainty of 4\%.

The tracking uncertainty includes two components. 
The first is the difference in track reconstruction
efficiency between data and simulation,
estimated with a tag and probe method~\cite{LHCb-PUB-2011-025} 
of $\jpsi\rightarrow\mu^+\mu^-$ decays, which is found to be negligible.
The second is due to the 2\% uncertainty on the effect from hadronic
interactions assumed in the detector simulation.

The uncertainty due to the choice of the $(p_{\rm T},\eta)$ binning is found 
to be negligible. Combining all systematic uncertainties in
quadrature, we obtain $R_{c/u} = \BcCSResult$
for $\Bc$ and $\Bu$ mesons with transverse momenta 
$p_{\rm T}>4$ GeV/$c$ 
and pseudorapidities $2.5<\eta<4.5$.

For the mass measurement, different selection criteria are applied. 
All events are used regardless of the trigger line.
The fiducial region requirement is also removed. 
Only candidates with a good measured mass uncertainty ($<20$ \mevcc) 
are used, and a loose particle identification requirement on the pion 
of the $\bcjpsipi$ decay is 
introduced to remove the small contamination from 
$B_c^+\to \jpsi K^+$ decays.

The alignment of the tracking system and the calibration of the
momentum scale are performed
using a sample of $\jpsi \to \mu^+\mu^-$ decays in periods
corresponding to different running conditions, as described 
in Ref.~\cite{Aaij:2011ep}.
The validity of the calibrated momentum scale has
been checked using samples of $\KS \to\pi^+\pi^-$ 
and $\Upsilon \to \mu^+\mu^-$ decays. 
In all cases, the effect of the final state radiation, 
which cause the fitted masses to be underestimated,
is taken into account. 
The difference between 
the correction factors determined using the $\jpsi$ and 
$\Upsilon$ resonances, 0.06\%, 
is taken as the systematic uncertainty. 

\begin{figure}
\centering
\includegraphics[width=0.495\textwidth]{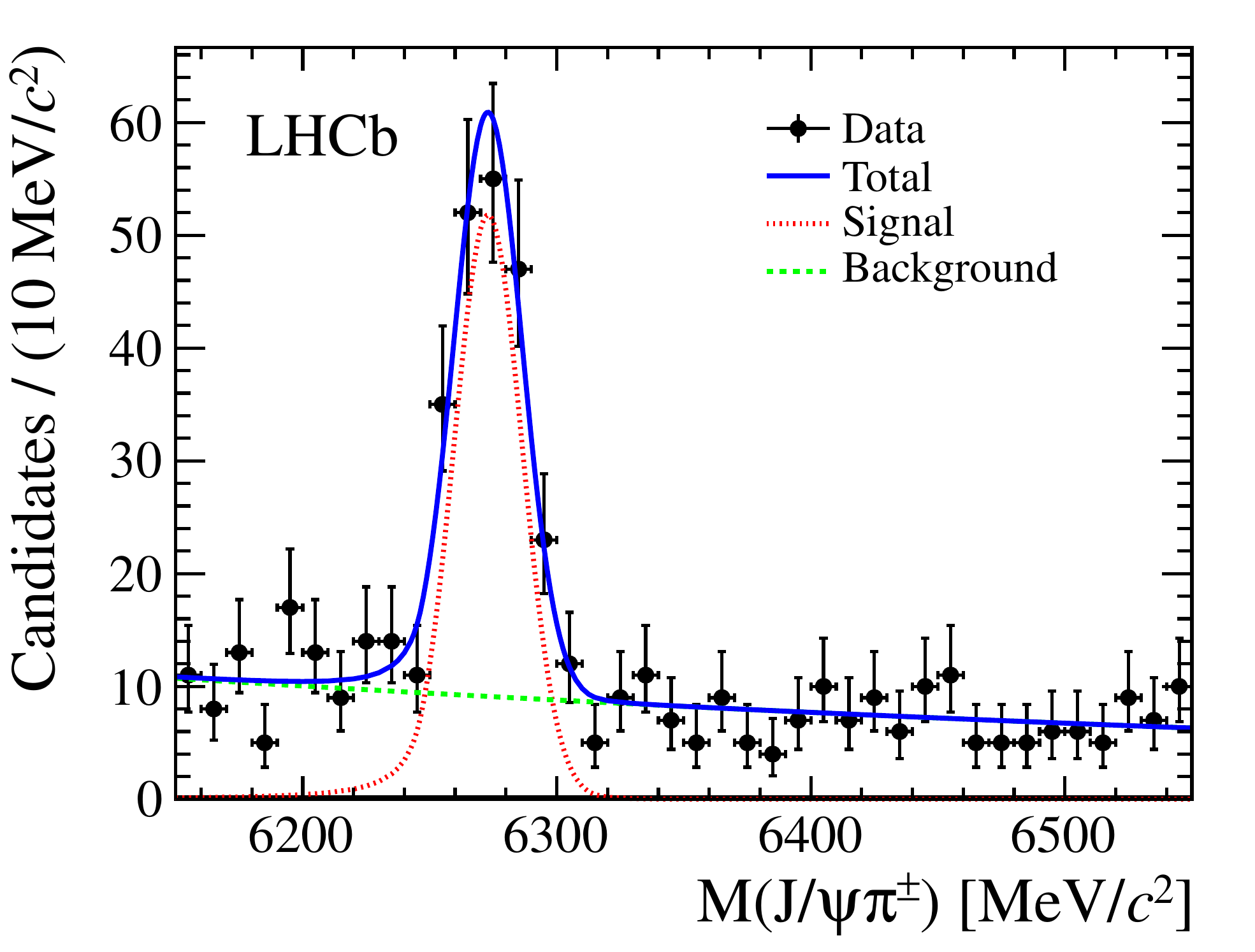}
\caption{Invariant mass distribution of $\bcjpsipi$ decays, used in
  the mass measurement. 
  The fit to the data is superimposed.} 
\label{fig:BcMass}
\end{figure}

The $\Bc$ mass is determined with an extended unbinned maximum
likelihood fit to the invariant mass distribution of the selected
$\bcjpsipi$ candidates. 
The mass difference $M(\Bc)-M(\Bu)$ is obtained by fitting
the invariant mass distributions of the selected
$\bcjpsipi$ and $\bpjpsik$ candidates simultaneously. 
The fit model is 
the same as in the production cross-section ratio measurement. 
Figure~\ref{fig:BcMass} shows the invariant mass distribution for
$\bcjpsipi$. 
The $\Bc$ mass is determined to be $6273.0 \pm 1.3\,\mevcc$,
with a resolution of $13.4 \pm 1.1\, \mevcc$, 
and the mass difference $M(\Bc)-M(\Bu)$ is $994.3 \pm 1.3\,\mevcc$.
The uncertainties are statistical only.

The mass measurement is affected by the systematic uncertainties due 
to the invariant mass model, momentum scale calibration, 
detector description and alignment.
To evaluate the systematic
uncertainty, the complete analysis, including the track
fit and the momentum scale calibration when needed, is repeated. 
The parameters to which the mass measurement is sensitive are varied 
within their uncertainties. The changes in the central values
of the masses obtained from the fits 
relative to the nominal results are then assigned as systematic 
uncertainties.

Table~\ref{tab:BcMassDiffSysSum} summarizes the systematic
uncertainties assigned to the measured $\Bc$ mass and mass
difference $\Delta M = M(\Bc)-M(\Bu)$.
The main source is the uncertainty in the momentum scale calibration. 
After the calibration procedure a residual $\pm 0.06$\% variation of the
momentum scale remains as a function of 
the particle pseudorapidity $\eta$.
The impact of this variation is evaluated by
parameterizing the momentum scale as a function of $\eta$.
The amount of material traversed by a particle in the tracking system 
is known to 10\% accuracy, the magnitude of the energy loss
correction in the reconstruction is therefore varied by 10\%.
To quantify the effects due to the alignment uncertainty, 
the horizontal and vertical slopes of the tracks close to
the interaction region, which are determined by measurements in the
vertex detector, are changed by $\pm$0.1\%, corresponding to the
estimated precision of
the length scale along the beam axis~\cite{Aaij:2011qx}.
To test the relative alignment of different sub-detectors, 
the analysis is repeated ignoring the hits of the tracking
station between the vertex detector and the magnet.
Other uncertainties arise from the signal and background line shapes. 
The bias due to the final state radiation is studied using a simulation
based on \textsc{Photos}~\cite{Golonka:2005pn}.
The mass returned by the fit model is found to be underestimated
by $0.7 \pm 0.1\,\mevcc$ for the $\Bc$ meson, 
and by $0.4 \pm 0.1\,\mevcc$ for the $\Bu$ meson.
The mass and mass difference are corrected accordingly, and
the uncertainties are propagated.
The effects of the background shape are evaluated by using a constant 
or a first-order polynomial function instead of the nominal exponential function.
The stability of the measured $\Bc$ mass is studied by dividing the
data samples according to the polarity of the spectrometer magnet and 
the pion charge. The measured $\Bc$ masses are consistent 
with the nominal result within
the statistical uncertainties.

\begin{table}
  \centering
  \caption{Systematic uncertainties (in $\mevcc$) of the $B_c^+$ mass
    and mass difference $\Delta M = M(\Bc)-M(\Bu)$.} 
  \label{tab:BcMassDiffSysSum}
  \begin{tabular}{ l c c}
    \hline
      Source of uncertainty & $M(B_c^+)$ & $\Delta M$  \\  

\hline
Mass fitting:         &   & \\
~ -- Signal model          &  $ 0.1$  & $ 0.1$ \\  
~ -- Background model  &   $ 0.3$ & $ 0.2$ \\

Momentum scale:    &   & \\
~ -- Average momentum scale &  $ 1.4$ & $ 0.5$ \\
~ -- $\eta$ dependence &  $ 0.3$ & $ 0.1$ \\

Detector description:    &   & \\
~ -- Energy loss  correction &  $ 0.1$ & -\\

Detector alignment:    &  & \\
~ -- Vertex detector (track slopes) & $ 0.1$ & - \\
~ -- Tracking stations &  $ 0.6$ & $ 0.3$ \\
\hline
Quadratic sum &   $ 1.6$ & $ 0.6 $ \\ 

\hline
\end{tabular}
\end{table}

In conclusion, using 0.37 fb$^{-1}$ of data collected
in $pp$ collisions at $\sqrt{s}=7$ TeV by the LHCb experiment, 
the ratio of the production cross-section times branching fraction of 
$\Bc \to \jpsi \pi^+$ 
relative to that for 
$\Bu \to \jpsi K^+$ is measured to be $R_{c/u}=\BcCSResult$
for $\Bc$ and $\Bu$ mesons with transverse momenta $p_{\rm T}>4$ GeV/$c$ 
and pseudorapidities $2.5<\eta<4.5$.
Given the large theoretical uncertainties on both production
and branching fractions of the $\Bc$ meson, 
more precise theoretical predictions are 
required to make a direct comparison with our result. 
The $B_c^+$ mass is measured to be $\BcMassResult$.
The measured mass difference with respect to the $\Bu$ meson 
is $M(B_c^+)-M(\Bu) = \BcMassDiffResult$.
Taking the world average $\Bu$ mass~\cite{PDG2012}, 
we obtain
$M(\Bc)=6273.9 \pm 1.3\,({\rm stat.}) \pm 0.6\,({\rm syst.})\,\mevcc$,
which has a smaller systematic uncertainty. 
The measured $\Bc$ mass is in agreement with previous
measurements~\cite{Aaltonen:2007gv, Abazov:2008kv} and a recent prediction
given by the lattice QCD calculation, $6278(6)(4) \mevcc$~\cite{Chiu:2007bc}.
These results represent the most precise determinations of these
quantities to date.

\section*{Acknowledgements}

\noindent We express our gratitude to our colleagues in the CERN accelerator
departments for the excellent performance of the LHC. We thank the
technical and administrative staff at CERN and at the LHCb institutes,
and acknowledge support from the National Agencies: CAPES, CNPq,
FAPERJ and FINEP (Brazil); CERN; NSFC (China); CNRS/IN2P3 (France);
BMBF, DFG, HGF and MPG (Germany); SFI (Ireland); INFN (Italy); FOM and
NWO (The Netherlands); SCSR (Poland); ANCS (Romania); MinES of Russia and
Rosatom (Russia); MICINN, XuntaGal and GENCAT (Spain); SNSF and SER
(Switzerland); NAS Ukraine (Ukraine); STFC (United Kingdom); NSF
(USA). We also acknowledge the support received from the ERC under FP7
and the Region Auvergne.

\addcontentsline{toc}{section}{References}
\bibliographystyle{LHCb}
\bibliography{main}

\end{document}